\documentclass{Interspeech}
\usepackage[table,xcdraw]{xcolor}
\usepackage{multirow}
\usepackage{cite}
\usepackage{hyperref}
\usepackage{booktabs}
\usepackage{siunitx}

\providecommand{\NZ}[1]{\textcolor{blue}{[{\bf #1}]}}

\interspeechcameraready % comment when submit

\title{PartialEdit: Identifying Partial Deepfakes in the Era of Neural Speech Editing 
\thanks{* Equal Contribution} % commented for double-blind
}

\author[affiliation={1}]{You}{Zhang*}
\author[affiliation={1}]{Baotong}{Tian*}
\author[affiliation={2}]{Lin}{Zhang}
\author[affiliation={1}]{Zhiyao}{Duan}

\affiliation{Audio Information Research Lab, University of Rochester}{Rochester}{USA}
\affiliation{Speech@FIT, Brno University of Technology}{Brno}{Czechia}
\email{\{you.zhang, baotong.tian, zhiyao.duan\}@rochester.edu, zlin@ieee.org}


\usepackage{comment}

\begin{document}

\maketitle

\begin{abstract}    
    Neural speech editing enables seamless partial edits to speech utterances, allowing modifications to selected content while preserving the rest of the audio unchanged. This useful technique, however, also poses new risks of deepfakes. To encourage research on detecting such partially edited deepfake speech, we introduce PartialEdit, a deepfake speech dataset curated using advanced neural editing techniques. We explore both detection and localization tasks on PartialEdit. Our experiments reveal that models trained on the existing PartialSpoof dataset fail to detect partially edited speech generated by neural speech editing models. As recent speech editing models almost all involve neural audio codecs, we also provide insights into the artifacts the model learned on detecting these deepfakes. 
    % These findings motivate our argument for a clearer definition of deepfake as content-generated speech segments. 
    % We hope that our proposed dataset curation process, experiments, and perspectives contribute to the development of partial deepfake detection research. 
    Further information about the PartialEdit dataset and audio samples can be found on the project page: \url{https://yzyouzhang.com/PartialEdit/index.html}.
    %provides a foundation for defending against partially edited speech in the era of neural speech editing. 
    % \lz{More details can be found at the project page url{https://yzyouzhang.com/PartialEdit/index.html}}
    % We plan to release the dataset upon publication.
\end{abstract}

\section{Introduction}
Recent advances in text-to-speech (TTS) and voice-conversion (VC) technologies have enabled the generation of audio that is virtually indistinguishable from genuine human speech~\cite{tan2021survey, ju2024naturalspeech}. The risk of their misuse by attackers to spread misinformation or attack security systems has grown significantly. Hence, deepfake detection has become an important area of research~\cite{li2025survey}.

Most existing work on speech deepfake detection targets cases in which entire utterances are synthesized by TTS or VC systems. 
However, in recent years, speech editing algorithms have been emerging~\cite{kassmann2024speech}, where users can generate audio in high quality by \textit{partially} modifying existing speech, rather than generating from scratch~\cite{jin2017voco, morrison2021context, Editspeech, wang2022campnet, peng2024voicecraft, wang2024ssr}. 
% As such, we argue that the exploration of the \emph{partial deepfake} scenarios, particularly with speech editing techniques, remains limited. 
Although partial deepfakes have been discussed in some literature~\cite{Zhang2021, zhang2022partialspoof, yi2023add}, they mainly focused on scenarios in which modified speech segments were generated using vocoder-based TTS and VC methods, then spliced back into the original utterance using basic signal processing techniques. 
However, the effectiveness of existing partial deepfake detection systems~\cite{zhang2022partialspoof, zhong2024enhancing} remains unverified against advanced speech editing techniques, where edited regions are generated through in-context learning, making them potentially more difficult to detect.

Additionally, recent speech generation models have transitioned to a neural codec-based paradigm, employing a flexible end-to-end generation pipeline, and are evolving rapidly. Unlike traditional vocoder-based approaches, modern models leverage neural audio codecs to represent speech as discrete tokens, which audio language models can process. This enables seamless speech editing through techniques such as prompting and infilling, making speech editing more natural and contextually coherent while preserving natural prosody and speaker traits. Although prior studies~\cite{wu24p_interspeech, xie2025codecfake, du2025codecfake} explored deepfake detection in neural codec-based speech generation, they overlooked speech editing models, which modify real recordings rather than generate fully synthetic speech. Since real-world misuse would often involve editing bona fide speech, our study addresses the unexplored challenge of the detection and localization of partially edited deepfakes in the era of neural speech editing.

\begin{figure}[!t]
	\centerline{\includegraphics[width=0.99\linewidth]{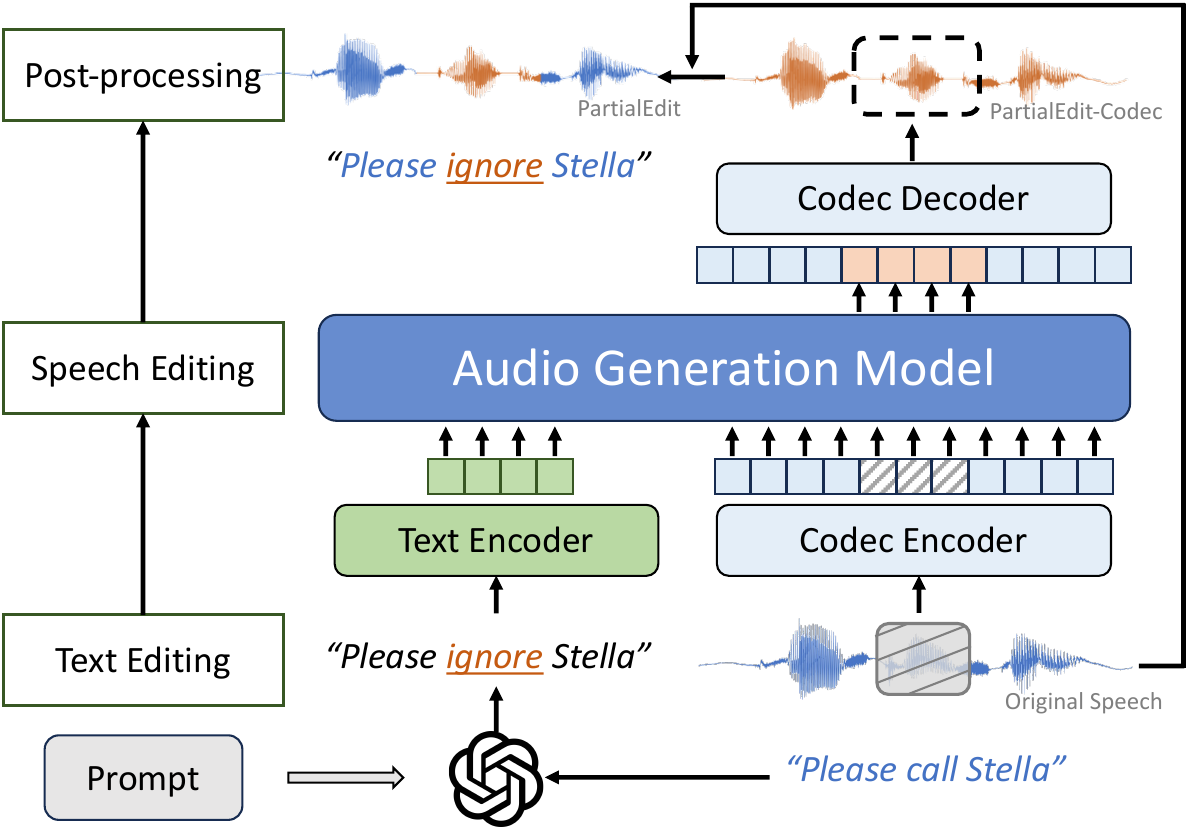}}
  \vspace{-1pt}
\caption{PartialEdit curation process, illustrated using the utterance \texttt{p226\_001} from the VCTK dataset. The original speech is modified to produce ``Please ignore Stella,''  where the speech segment corresponding to ``ignore'' is synthesized, making it the \textbf{partial deepfake} within the newly edited utterance.}
 \vspace{-4pt}
\label{fig:curation}
\end{figure}

We introduce \textbf{PartialEdit}, a new partial deepfake dataset that involves neural speech editing, with its curation process illustrated in Figure~\ref{fig:curation}. PartialEdit is built on various modern speech editing models, including VoiceCraft~\cite{peng2024voicecraft}, SSR-Speech~\cite{wang2024ssr}, Audiobox-Speech, and Audiobox~\cite{vyas2023audiobox}. A previous study~\cite{huang2024detecting} is close to ours, but they only considered Voicebox (a vocoder-based method that preceded Audiobox) and did not handle more advanced neural codec-based speech editing methods. We conduct partial deepfake detection and localization on PartialEdit and find that detectors trained on existing partially spoofed audio fail to generalize to PartialEdit. We also perform deepfake localization comparisons across different speech editing methods and find that audio partially edited by VoiceCraft and SSR-Speech is harder to detect compared to Audiobox-Speech and Audiobox. Moreover, we discuss the impact of post-processing where stitching artifacts are introduced under codec-processed artifacts, and argue for a clearer definition of deepfake regions that focuses solely on content-edited segments, regardless of codec-introduced artifacts.
% unedited parts back to audio and from neural codec resynthesis. 
Our results provide new insights into the detection and localization of partially deepfake audio in the era of neural speech editing. 

% \input{0_introduction_lin}
% \input{0_introduction_baotong}
%%%%%%%%%%%%%%%%%%%%%%%

\section{PartialEdit Dataset}
% In this section, we first briefly overview the speech editing models in PartialEdit, followed by a detailed description of the dataset curation process, and conclude with dataset statistics.
In this section, we briefly overview the neural speech editing models used in curating our PartialEdit dataset, describe the dataset curation process, and present dataset statistics.

% \subsection{Codec-based neural speech editing}
% Figure for creating database. 
% \NZ{This would be more informative than figure 1. Demonstrating we have four methods and two fashion (Voicecraft/SSR; Audiobox-speech/joint)}

% voicebox vs. audiobox
% ref: https://ai.meta.com/blog/audiobox-generating-audio-voice-natural-language-prompts/
% This work is heavily built upon the training objective and model architecture of Voicebox (Le et al., 2023), and the self-supervised objective of SpeechFlow (Liu et al., 2023a). 
% Audiobox inherits Voicebox’s guided audio generation training objective and flow-matching modeling method to allow for audio infilling. With infilling, users can also use the model to polish sound effects (adding different thunder sounds into a raining soundscape, for example).
% Sound editing with generative infilling: Users can crop an audio segment and regenerate it with Audiobox. By providing a text description, Audiobox can insert sound effects like “a dog barking” into an audio clip of the sound of rain.

\subsection{Neural speech editing models used in PartialEdit}\label{sec:edit_model}

% \NZ{An overview of speech editing methods in general.} Codec-based LM, shown in the dashed box in Figure
We adapt the following codec-based speech generation models designed for or capable of speech editing: VoiceCraft~\cite{peng2024voicecraft}, SSR-Speech \cite{wang2024ssr}, Audiobox-Speech and Audiobox \cite{vyas2023audiobox}. All of them utilize Encodec~\cite{defossez2022high_encodec} but with different configurations. 
% The Encodec model used in both VoiceCraft (E1) and SSR-Speech (E2) operates at a frame rate of 50 Hz for audio recorded at a 16 kHz sampling rate. It is trained on the GigaSpeech dataset using the Audiocraft repository\footnote{\href{https://github.com/facebookresearch/audiocraft}{https://github.com/facebookresearch/audiocraft}}. In contrast, the Encodec model used in Audiobox operates at a 24 kHz sampling rate. 
To maintain consistency, we downsample all samples generated by Audiobox from 24 kHz to 16 kHz.
% In general, the audio language models take text tokens 

\textbf{VoiceCraft (E1)}~\cite{peng2024voicecraft} formulates sequence infilling (for speech editing) by rearranging tokens from the neural audio codec.
The original speech is first converted to discrete codec tokens by the Encodec~\cite{defossez2022high_encodec} encoder. A subset of tokens is masked and shifted to the end of the sequence. The target transcript, together with these processed tokens, is fed into a decoder-only Transformer, which autoregressively predicts the masked tokens.
Surrounding frames are slightly modified to ensure smooth transitions, and the predicted tokens are rearranged and decoded back into audio by the Encodec decoder.
% The masked spans are determined by comparing the original and modified transcripts, aligning the original transcript with the audio codec tokens by WhisperX \cite{bain2022whisperx}. The autoregressively generated tokens are then seamlessly spliced back into their respective locations, and the entire sequence is mapped back to audio using the Encodec decoder. 
% While the original VoiceCraft is limited to editing a single span at a time, we extended its capability to edit multiple spans by iteratively processing one span at a time, following the prescribed editing approach.

\textbf{SSR-Speech (E2)}~\cite{wang2024ssr} is built on VoiceCraft with key improvements. 
In particular, SSR-Speech can automatically detect the type of edit, whether insertion, deletion, or substitution, and apply the appropriate modifications accordingly, whereas VoiceCraft only allows one edit type provided by the user.
% In particular, a specialized Watermark Encodec, designed specifically for speech editing, is incorporated into SSR-Speech. However, during dataset curation, we did not apply this mechanism, as its impact on audio quality compared to the original Codec remains unclear, and its inclusion would introduce an additional watermark detection task.

% The original transcription of VCTK is not used by VoiceCraft and SSR as we applied 

\noindent Note that for both VoiceCraft and SSR-Speech, instead of using the original transcripts of VCTK, we follow the original structure of SSR-Speech to apply WhisperX\footnote{\href{https://github.com/m-bain/whisperX}{https://github.com/m-bain/whisperX}} \cite{bain2022whisperx} to produce transcriptions as well as word-level alignment. 
% Whereas in Audiobox,  [Alignment difference]

\textbf{Audiobox-Speech (E3)}~\cite{vyas2023audiobox} is an Encodec-based speech generation model based on flow-matching. 
% audio language model 
It fine-tunes the self-supervised generative pre-training foundation model for in-context TTS using transcribed speech. For speech editing, the original transcript and the speech are aligned 
% the process begins by aligning the original transcript with the original speech 
using a forced aligner of character-level char-units~\cite{seamless2023}. Given the target transcript, a new alignment is made with a preset masked duration of the edited region. The duration of the edited region is sampled using a pre-trained flow-matching duration model and then serves as conditional input for the audio flow. 

\textbf{Audiobox (E4)}~\cite{vyas2023audiobox} is a unified model capable of generating both speech and general audio, conditioned on text descriptions or audio examples. % It encompasses the functionalities of both Audiobox-Speech and Audiobox-Sound while also allowing speech to be re-stylized within a new acoustic environment. In this work, we utilize Audiobox specifically for speech editing. 
The generation process remains the same as in Audiobox-Speech, and it can be considered a variant with different parameter weights due to comprehensive training objectives, but its additional capabilities for generating sound are not activated for this neural speech editing application.

\subsection{PartialEdit curation process}\label{sec:data_creation}

We use the VCTK \cite{Veaux2016CSTRToolkit} dataset as the source of bona fide speech, consistent with several widely used deepfake datasets~\cite{Wang2020data_asvspoof2019, wu24p_interspeech}, and the partial deepfake dataset PartialSpoof~\cite{zhang2022partialspoof}.
To generate high-quality, partially edited deepfake speech, we follow a three-step process
% : text editing, speech editing, and post-processing, 
as illustrated in Figure~\ref{fig:curation}: 

\noindent\textbf{Step 1: Text editing}. To ensure naturalness after text modifications, we adopt an approach inspired by LlamaPartialSpoof~\cite{luong2024llamapartialspoof}. Specifically, we iteratively input each transcript from VCTK~\cite{Veaux2016CSTRToolkit} into GPT-4o-mini, prompting it to modify one word\footnote{Although the prompt instructs GPT to modify only a single word, maintaining sentence naturalness occasionally requires modifying two words, such as changing ``included in'' to ``excluded from.''} while preserving grammatical correctness and fluency. 
% This ensures that the modified sentence remains coherent.
% while not relying on manual text crafting.

\noindent\textbf{Step 2: Neural speech editing}. 
% As illustrated in Figure~\ref{fig:curation}, the modified transcripts from Step 1 are processed by a text encoder into text tokens. We then align the original transcript with the audio and decide on the masked regions for edits by comparing the original and modified transcripts. The original speech is then fed into a codec encoder to convert it into discrete tokens, where a portion is masked. The audio generation models use these inputs to generate tokens for the masked regions, retaining the tokens for the unedited portions. Then, the newly constructed tokens are converted back to audio through the codec decoder. 
As shown in Figure~\ref{fig:curation}, our pipeline first encodes the modified transcripts from Step 1 into text tokens by a text encoder. Next, we align the original speech to its transcript and compare word-level differences to identify edit regions.
The speech waveform is then converted to discrete tokens via a neural codec encoder. Tokens corresponding to designated edit regions are masked, while the audio generation model predicts the masked tokens conditioned on both text tokens and unmasked speech context. We preserve the unmasked tokens to maintain consistency in content-unedited segments. Finally, the modified token sequence—comprising both the newly predicted tokens and the retained original tokens—is decoded into an audio waveform using the neural codec decoder, producing the edited speech output.
% Then, we pass the original speech through a codec encoder, along with both the original and modified transcripts, into the audio generation models. These models generate speech tokens corresponding to the masked regions while retaining tokens for the unedited portions of the original utterance. 
% In case of multi-span editing brought by either the GPT modification or transcription problems, we perform edit once at a time for E1, E3 and E4, whereas E2 natively supports multi-span editing. 
% Note that content-unedited parts still inherit neural codec artifacts after encoding and decoding, as explored in CodecFake~\cite{wu24p_interspeech}. 
% To mitigate these artifacts, we apply an additional post-processing step.
% Even unedited parts inherit neural codec artifacts after encoding/decoding

\noindent\textbf{Step 3: Post-processing}. In speech editing models, although only the content-edited regions are intentionally generated, the entire output audio, including content-unedited regions, undergoes neural codec processing. To avoid additional artifacts introduced by neural audio codecs, we introduce a cut-and-paste post-processing step rather than directly using the output from speech editing models.
% to stitch content-eidted part back into the original bona fide audio. 
Specifically, with alignment information derived from the speech editing model, we extract the edited parts from the generated speech and cut and paste them back into the original audio. This cut-and-paste operation, also adopted in PartialSpoof~\cite{zhang2022partialspoof}, ensures that the original bona fide speech is preserved in the resulting output.
Detailed discussions on the artifacts introduced by neural codecs are in Section~\ref{sec:codec_real_fake}.
% We assume that the artifacts introduced by this operation are negligible, given that only limited word parts are manipulated.
% according to the alignment, ensuring temporal consistency.

% \subsection{Different types}

\begin{figure}[!t]
	\centerline{\includegraphics[width=0.94\linewidth]{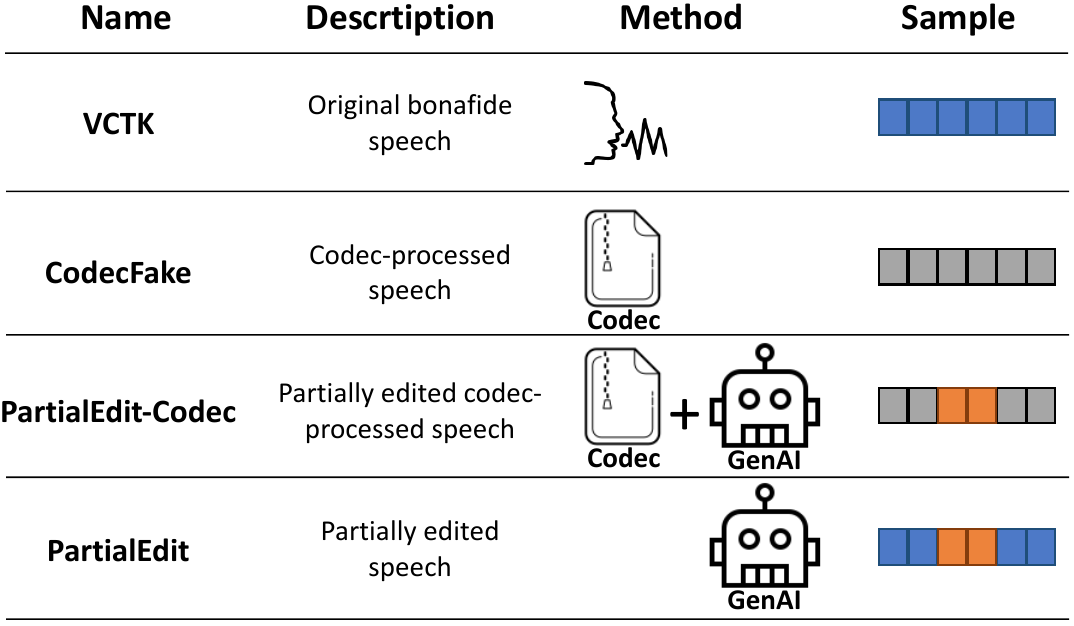}}
  \vspace{-1pt}
% \caption{Overview of different speech processing cases considered in this study. The bottom two, PartialEdit-Codec and PartialEdit, constitute our newly curated PartialEdit dataset. }
\caption{Overview of different cases considered in this study. Our analysis decouples codec processing from generation.}\label{fig:case_definition}
 \vspace{-3pt}
\end{figure}

As shown in Figure \ref{fig:curation}, we construct \textbf{PartialEdit} using audio produced from the final post-processing step, where only the content-edited regions are generated by the speech editing model.
% , aligning with previous investigations on partial spoof~\cite{zhang2022partialspoof}. 
We also retain an intermediate version of the dataset, \textbf{PartialEdit-Codec}, which contains deepfake speech from Step~2 without post-processing. The key distinction between PartialEdit and PartialEdit-Codec is whether the content-unedited parts have passed through the neural codec. In particular, although the content for the content-unedited region remains unchanged in both cases, the PartialEdit-Codec version introduces additional neural codec processing. We direct readers to Figure~\ref{fig:case_definition} for an intuitive comparison by visualization.
% Thus, unless otherwise specified, we conduct most of our experiments 
% Further discussion on the relationship between PartialEdit and PartialEdit-Codec can be found in Section \ref{sec:codec_real_fake}.

% One might argue that speech editing is already completed after Step 2. However, we investigate the strongest attack scenario, where an adversary aims to make the manipulated speech as realistic as possible. In this case, attackers would want the fake words to be hidden seamlessly within genuine speech, ensuring that the edited information is convincingly spread while the rest of the utterance remains natural.

% \begin{enumerate}
%     \item Bona fide (VCTK) 
%     \item CodecFake
%     \item \textcolor{red}{PartialEdit-codec}
%     \item \textcolor{red}{PartialEdit}
% \end{enumerate}

\subsection{Dataset statistics information}\label{sec:data_statistic}

We maintain the same utterance across all subsets generated by different speech editing models. After removing unsuccessfully edited utterances from all subsets,
% we delete the corresponding utterance across all subsets. 
% This results in 
the dataset consists of 108 speakers and 43,358 partially edited utterances from each speech editing model. Following the speaker setup of ASVspoof2019 \cite{Wang2020data_asvspoof2019} and PartialSpoof \cite{zhang2022partialspoof}, we split speakers and utterances into three disjoint partitions: 20 speakers (8,258 utterances), 20 speakers (7,915 utterances), and 68 speakers (27,185 utterances) for training, validation, and evaluation sets, respectively.
% The speaker splits follow ASVspoof~\cite{Wang2020data_asvspoof2019} and PartialSpoof~\cite{zhang2022partialspoof}.
% \NZ{Let's say this for now but organize the dataset before camera ready. Then we only need to report the average duration and MOS. I don't think we need to report separately for train/dev/eval. }
We applied a pretrained DistillMOS~\cite{stahl2025distillationpruningscalableselfsupervised} model to estimate the mean opinion score (MOS) as a measure of the naturalness of speech generated by each editing model. 

% Duration and predicted MOS for 
Table~\ref{tab:data_info} shows the duration and predicted MOS for both PartialEdit and PartialEdit-Codec subsets.
Consistently, PartialEdit achieves a higher MOS than PartialEdit-Codec across all speech editing models, suggesting that the artifacts introduced by cutting and pasting the edited region back into the original speech are less noticeable than those introduced by neural codec processing.
% highlighting the useful of the additional post-processing. In this processing, the artifact introduced by cutting modiffied region back to the may less than those introduced by codec processing.
% all subsets, 
% highlighting the naturalness of the partially edited speech. 
The overall MOS of PartialEdit is similar to that of bona fide speech from VCTK. 
% Though the MOS score imperfectly aligns with human listening, we suspect the partially edited deepfake speech is not distinguishable from humans with high-quality comparable to bona fide speech.
This indicates that partially edited deepfake speech in our curated PartialEdit dataset achieves perceptual closeness to bona fide speech; This matches what we subjectively noticed while curating the dataset.

\begin{table}[!t]
\centering
% \caption{Duration (h) and MOS of PartialEdit and PartialEdit-Codec. Duration is shown as Train/Dev/Eval, and MOS is shown as PartialEdit (PartialEdit-Codec).}
\caption{Duration (hours) and predicted mean opinion score (MOS) for PartialEdit and (PartialEdit-Codec). Duration report as train/dev/eval splits and shares between both versions.}
\label{tab:data_info}
\scalebox{0.83}{
\begin{tabular}{lcc}
\toprule
\textbf{Subset} & \textbf{Duration (h)} & \textbf{MOS} \\
\midrule
VCTK~\cite{Veaux2016CSTRToolkit}  & 7.80 / 8.18 / 25.13 & 3.88$\pm$0.28                   \\
% PartialSpoof-Fake~\cite{zhang2022partialspoof}  &  &    3.86$\pm$0.50               \\
% CodecFake-E~\cite{wu24p_interspeech} & 8.20/8.07/27.70 & 2.88$\pm$0.27                   \\
% \midrule
% \multicolumn{3}{c}{\textit{PartialEdit (PartialEdit-Codec)}}\\
\midrule
VoiceCraft \hfill (E1)      & 8.28 / 8.06 / 27.79 & 3.80$\pm$0.32 (3.60$\pm$0.38) \\
SSR-Speech \hfill (E2)      & 7.82 / 7.64 / 26.26 & 3.83$\pm$0.30 (3.71$\pm$0.34) \\
Audiobox-Speech \hfill (E3) & 7.94 / 7.96 / 25.69 & 3.90$\pm$0.32 (3.53$\pm$0.32) \\
Audiobox \hfill (E4)        & 8.14 / 7.96 / 26.44 & 3.90$\pm$0.33 (3.54$\pm$0.32) \\
\bottomrule
\end{tabular}
}
% \vspace{-0pt}
\end{table}

% \section{PartialEdit Detection}
\section{Detection on partially edited deepfakes
% \\ Detection on PartialEdits
}
\label{sec:detection}
% \NZ{In this section, we consider CodecFake and PartialEdit-Codec as all fake. Only blue parts in Figure~\ref{fig:case_definition} is real. The discussion on whether the grey part should be considered real is moved to the next section.}

Partial deepfake detection involves two complementary tasks: utterance-level detection and segment-level localization. In this section, we focus on utterance-level detection---determining whether an entire speech sample is bona fide or contains partial edits. We will then discuss localization in Section \ref{sec:localization}.
% \nz{We include VCTK in experiments for bona fide audio.}

\subsection{Experimental setup}
% \subsection{Partial audio deepfake detection methods}
% survey~\cite{li2025survey}
% \textbf{Tasks}
% detection
% localization
% \textbf{Models}
% SoTA in each task
% Utterance-level methods:
% SSL+AASIST~\cite{tak22_odyssey}, as it serves as baselines for various benchmarks

\noindent\textbf{Datasets}. 
Besides PartialEdit and PartialEdit-Codec datasets (where all speech generation models E1-E4 are included), we also incorporate PartialSpoof~\cite{zhang2022partialspoof} dataset in this study. PartialSpoof~\cite{zhang2022partialspoof} is a widely used dataset for partial deepfake detection, where random speech segments from real utterances are replaced with deepfake speech.
% and concatenated to create partial deepfakes. 
% The fully deepfake utterances are sourced from ASVspoof2019, which employs traditional TTS and VC methods for fake speech generation. 
Both PartialSpoof and our PartialEdit share the VCTK~\cite{Veaux2016CSTRToolkit} corpus as the bona fide source. 

\noindent\textbf{Model configuration}. We select XLSR-SLS~\cite{zhang2024audio} to perform deepfake detection, as it achieves top performance on various audio deepfake detection benchmarks~\cite{speecharena-df-leaderboard, zhang2024svdd}. It adopts a large-scale self-supervised learning (SSL) representation XLS-R\footnote{\href{https://dl.fbaipublicfiles.com/fairseq/wav2vec/xlsr2_300m.pt}{https://dl.fbaipublicfiles.com/fairseq/wav2vec/xlsr2\_300m.pt}}~\cite{babu22_interspeech} as the front-end and incorporates a sensitive layer selection (SLS) module as the back-end. 
% 
% model in the PartialEdit dataset
% 
% \noindent\textbf{Training Details}.
We use the same set of hyperparameters for training following~\cite{zhang2024audio}. We train each model respectively for 10 epochs and set 3 for early stopping.

\noindent\textbf{Evaluation metrics}.
We use the equal error rate (EER) to present the performance of deepfake detection.
\subsection{Utterance-level deepfake detection}
\label{ssec:utter}
% \NZ{The goal of this section is to tell the audience that our dataset is useful. The audio language model does pose challenges to the partial deepfake detection problem. }

Our results are presented in Table~\ref{tab:cross_datasets}. The model trained on PartialSpoof (Row I) fails to generalize to our PartialEdit dataset, as indicated by the high EERs in Columns II and III. This suggests that our dataset presents new challenges for partial deepfake detection.
Not surprisingly, training on PartialEdit-Codec (Row II) or on PartialEdit (Row III) shows prominently better performance on their same test data. Interestingly, training on PartialEdit generalizes well to PartialEdit-Codec (Row III, Column II) but not the other way around (Row II, Column III). This seems to suggest that artifacts presented in PartialEdit-Codec (mainly codec-related) are also presented in PartialEdit (codec-related and stitching-related), but not the other way around.
%significantly improves model performance on PartialEdit (2.14\% EER) and PartialEdit-Codec (0.41\% EER); While the former is not surprising, the latter suggests that the majority of artifacts presented in PartialEdit-Codec  , demonstrating that speech editing-based partial deepfakes remain distinguishable and can benefit from targeted training. Notably, the model trained on PartialEdit generalizes well to PartialEdit-Codec (0.41\% EER) while staying well on itself (2.14\% EER).
% Interestingly, models trained on PartialEdit-Codec (row II) do not generalize well to PartialEdit-Stitch (column III, 27.59\% EER), suggesting that stitching with the original speech segments is more challenging for deepfake detection. 
% This is similar with previous findings where models trained on ASVspoof failed to generalize well to PartialSpoof~\cite{zhang2022partialspoof}, reinforcing the importance of partial deepfake detection.

Additionally, the EERs on PartialEdit (Column III) are consistently higher than those on PartialEdit-Codec (Column II), indicating that partial deepfakes where unedited segments remain identical to the original are more challenging for detection systems. This suggests that the artifacts introduced by post-processing with cutting and pasting are less detectable than those introduced by codec processing. This finding also aligns with their predicted MOS values discussed in Section \ref{sec:data_statistic}. 
% This highlights the importance of post-processing for attackers, as it increases the difficulty of detection.

% All models trained on codec-based deepfake datasets (rows II, III) generalize well to CodecFake-E (column II), emphasizing that neural speech editing presents a stronger challenge than CodecFake. 
However, neither of the models trained on our PartialEdit datasets (Rows II, III) generalizes to PartialSpoof (Column I), indicating that PartialSpoof represents a different paradigm compared to PartialEdit.
% This suggests that PartialEdit complements existing datasets, addressing previously unaccounted-for challenges in partial deepfake detection. 
% \lz{Mixing those two database together for training is left for future work.} \NZ{GPU running now.}
% \lz{...d to PartialEdit. 
% This suggests that when working with more advanced generation models, there is also a potential risk of overlooking conventional scenarios. Therefore, incorporating diverse training data is crucial.}
% \NZ{Train on PartialSpoof, Test on PartialEdit}
% Codec is challenging.
% Research has demonstrated that the model trained on conventional vocoder-generated spoofed data cannot work well on CodecFake~\cite{wu24p_interspeech}. We also do a similar task that when trained on PartialSpoof, it cannot work well on our PartialEdit.
% Demonstrate that the new speech editing methods pose challenges for existing partial deepfake audio detection methods.
% We use this condition to explore how generalization is affected when modeling technologies evolve.
% \NZ{Train on CodecFake-E, Test on PartialEdit}
% Partial is challenging.
% We take the E subset 
% This is to demonstrate that the Codec detection methods cannot generalize to partial detection. Similar to how models trained on ASVspoof2019 cannot generalize to PartialSpoof~\cite{zhang2022partialspoof}.
% \lz{PS+PE:} \NZ{Finish tomorrow morning}
% \lz{
When mixing PartialSpoof and PartialEdit to train the model (I+III), we observe promising results on all datasets, underscoring the need for anti-spoofing systems to tackle both cutting-edge deepfakes and conventional ones.
% This sugguests that although advanced generation models continue to emerge, producing higher-quality deepfakes, models trained solely on these advanced models may still fail to detect deepfakes generated by conventional technologies. 
% This highlights an important insight for the anti-spoofing community: while addressing deepfakes created by cutting-edge technologies, it remains crucial to ensure that detection systems also retain the ability to identify deepfakes generated by conventional methods. }

% (If we treat encodec as full fake, we cay say:). We use this case to investigate under the speech editing era, how detection becomes more difficult when only a portion of the audio is manipulated. 

\begin{table}[!t]
\caption{EERs (\%) of deepfake detection on PartialEdit and existing datasets. Rows correspond to training data for the model, while columns correspond to test data.
%Each row demonstrates one system trained on a specific setting, and each column demonstrates its test results on various datasets. 
The same applies to the following tables.
}\label{tab:cross_datasets}
\begin{minipage}{\linewidth}
\centering
\scalebox{0.96}{
% \centering
\begin{tabular}{l S[table-format=2.2] S[table-format=2.2] S[table-format=2.2] S[table-format=2.2]}
\toprule
{Train \textbackslash Test} & {I} & {II} & {III} \\ % & {IV} \\
\midrule
% PartialSpoof \hfill (I) &  2.55   & 12.95 & 23.72  \\ %  1.63&
% % II &CodecFake-E &  17.51 &  0.01 & 18.96  &  33.96 \\ \hline
% PartialEdit-Codec \hfill (II) & 14.54    &  0.13 & 27.59  \\ % & 0.04
% PartialEdit \hfill (III) &  23.06   & 0.41  & 2.14 \\ \hline  %& 0.19
% I + III & 3.00  & 0.64 & 2.61\\ % & 0.30

PartialSpoof \hfill (I) & \cellcolor[rgb]{0.98,0.98,0.98} \textcolor[rgb]{0.00,0.00,0.00}{~~2.55} & \cellcolor[rgb]{0.88,0.88,0.88} \textcolor[rgb]{0.00,0.00,0.00}{12.95} & \cellcolor[rgb]{0.79,0.79,0.79} \textcolor[rgb]{0.00,0.00,0.00}{23.72} \\ %  1.63 &
% II & CodecFake-E & \cellcolor[rgb]{0.84,0.84,0.84} \textcolor[rgb]{0.00,0.00,0.00}{17.51} & \cellcolor[rgb]{1.00,1.00,1.00} \textcolor[rgb]{0.00,0.00,0.00}{0.01} & \cellcolor[rgb]{0.83,0.83,0.83} \textcolor[rgb]{0.00,0.00,0.00}{18.96} & \cellcolor[rgb]{0.69,0.69,0.69} \textcolor[rgb]{0.00,0.00,0.00}{33.96} \\  \hline
PartialEdit-Codec \hfill (II) & \cellcolor[rgb]{0.87,0.87,0.87} \textcolor[rgb]{0.00,0.00,0.00}{14.54} & \cellcolor[rgb]{1.00,1.00,1.00} \textcolor[rgb]{0.00,0.00,0.00}{~~0.13} & \cellcolor[rgb]{0.75,0.75,0.75} \textcolor[rgb]{0.00,0.00,0.00}{27.59} \\ % &  0.04
PartialEdit \hfill (III) & \cellcolor[rgb]{0.79,0.79,0.79} \textcolor[rgb]{0.00,0.00,0.00}{23.06} & \cellcolor[rgb]{1.00,1.00,1.00} \textcolor[rgb]{0.00,0.00,0.00}{~~0.41} & \cellcolor[rgb]{0.98,0.98,0.98} \textcolor[rgb]{0.00,0.00,0.00}{~~2.14} \\  \hline  % &  0.19
I + III & \cellcolor[rgb]{0.97,0.97,0.97} \textcolor[rgb]{0.00,0.00,0.00}{~~3.00} & \cellcolor[rgb]{0.99,0.99,0.99} \textcolor[rgb]{0.00,0.00,0.00}{~~0.64} & \cellcolor[rgb]{0.98,0.98,0.98} \textcolor[rgb]{0.00,0.00,0.00}{~~2.61} \\ % &  0.30

% E2 & 6.50  & 3.57  & 9.17 & 9.33  \\
% E3 & 22.35 & 20.86 & 0.11 & 0.14 \\
% E4 & 16.32 & 15.32 & 0.17 & 0.11  \\

\bottomrule
\end{tabular}
}
\end{minipage}
\vspace{-2pt}

\end{table}

% \NZ{Train on PartialEdit-Codec, Test on PartialEdit}

% \NZ{Train on PartialEdit, Test on PartialEdit}
% Show that the model works better on this

% \NZ{Train on PartialEdit, Test on PartialSpoof}
% Maybe not working

% \section{PartialEdit Localization}
\section{\hspace{-3pt}Localization on partially edited deepfakes}\label{sec:localization}
This section focuses on the deepfake localization task, aiming to locate the edited regions within partially edited deepfakes.
% partially edited deepfakes. 
% This process is crucial for in-depth analysis and understanding the attackers' intent.

\subsection{Experimental setup}
% \noindent
\textbf{Datasets}. 
We conduct localization experiments on both the entire PartialEdit dataset and individual subsets generated by different speech editing models (E1-E4 in Table \ref{tab:data_info})
% (E1: VoiceCraft, E2: SSR-Speech, E3: Audiobox-Speech, E4: Audiobox) 
to examine how training on deepfakes generated by different speech editing methods affects the final localization performance.
% Since PartialEdit has been identified as both challenging and complementary to PartialSpoof in Section~\ref{ssec:utter}, we focus solely on PartialEdit for the deepfake localization task. This allows for a more in-depth comparison of partial deepfakes generated by different speech editing methods.

\noindent\textbf{Model configuration}. We adopt BAM~\cite{zhong2024enhancing} given its state-of-the-art performance on deepfake localization on the PartialSpoof dataset. 
% It employs a SSL model to extract speech presentations and leverages boundary enhancement with frame-wise attention to incorporate boundary information, guiding the segment-level discrimination between bona fide and spoof frames. 
% It first leverages an SSL model to extract robust speech features. Boundary enhancement and frame-wise attention then guide the segment-level discrimination between bona fide and spoofed frames. 
% 
% \noindent\textbf{Training Details}. 
Following the configuration in \cite{zhong2024enhancing}, we utilize WavLM-Large\footnote{\href{https://github.com/microsoft/unilm/tree/master/wavlm}{https://github.com/microsoft/unilm/tree/master/wavlm}}~\cite{WavLM} as the front-end and train the model at a 20 ms resolution. The training speech length is fixed at 4 seconds. We employ the Adam optimizer with an initial learning rate $10^{-5}$ that is then halved every 10 epochs. We also employ early stopping if the validation loss fails to reduce for 3 epochs. 

\noindent\textbf{Evaluation metrics}. We use frame-level EER with a 20 ms resolution to measure the performance of deepfake localization.

% \subsection{Segment-level Deepfake Localization}
% \subsection{Cross-generation method detection}
\subsection{Localization across different speech editing algorithms}
% This subsection discusses localization performance in speech editing, with a focus on the cross different speech editing algorithms. It aims
% We conduct Cross-speechediting experiments 
% to explore how PartialEdit speech generated by different speech editing technologies effect performance of spoof localization. Results are shown in Table \ref{tab:loc_cross_speechedit}.
% This subsection examines localization performance in speech editing, focusing on a comparison of different speech editing algorithms. We investigate how PartialEdit speech generated by various methods affects spoof localization performance. 

% \begin{table}[!thb]
% \caption{Cross-SpeechEditing experiments on Spoof Localization. All set includes [] as bona fide source.}\label{tab:loc_cross_speechedit}
% \centering
% \begin{tabular}{rccccc}
% \toprule
% Training \textbackslash Testing & E1 & E2 & E3 & E4 & PartialEdit\\
% \midrule
% E1 & 3.80/87.12  & 3.61/87.75  & 6.79/75.70 & 7.75/76.39 &\\
% E2 & 6.50/82.04  & 3.57/88.83  & 9.17/65.63 & 9.33/68.82 &\\
% E3 & 22.30/19.10 & 20.80/21.28 & 0.11/99.65 & 0.14/99.62 &\\
% E4 & 16.30/29.71 & 15.30/31.65 & 0.17/99.53 & 0.11/99.72 &\\
% % E12 (low)         &    &    &    &   & \\
% % E34 (low)         &    &    &    &   & \\
% PartialEdit             &    &    &    &   & \\
% \bottomrule
% \end{tabular}
% \end{table}

\begin{table}[!t]
\caption{Frame-level EERs (\%) of localization with cross-algo-rithm evaluation on different editing algorithms of PartialEdit.}\label{tab:loc_cross_speechedit}
\centering

\scalebox{0.95}{
\centering
% \begin{tabular}{rccccc}
\begin{tabular}{c S[table-format=2.2] S[table-format=2.2] S[table-format=2.2] S[table-format=2.2] S[table-format=2.2]}
\toprule
 {Train \textbackslash Test} & {E1}  & {E2}  & {E3}  & {E4}  & {PartialEdit} \\
\midrule
% E1 & 3.80  & 3.61  & 6.79 & 7.75 & 7.10 \\
% E2 & 6.50  & 3.57  & 9.17 & 9.33 & 9.51 \\
% E3 & 22.35 & 20.86 & 0.11 & 0.14 & 15.26 \\
% E4 & 16.32 & 15.32 & 0.17 & 0.11 &  11.77 \\
% % E12 (low)         &    &    &    &   & \\
% % E34 (low)         &    &    &    &   & \\
% \midrule
% PartialEdit & 4.07 & 3.30 & 0.18 & 0.16 & 2.77 \\
E1 & \cellcolor[rgb]{0.97,0.97,0.97} \textcolor[rgb]{0.00,0.00,0.00}{~~3.80} & \cellcolor[rgb]{0.97,0.97,0.97} \textcolor[rgb]{0.00,0.00,0.00}{~~3.61} & \cellcolor[rgb]{0.94,0.94,0.94} \textcolor[rgb]{0.00,0.00,0.00}{~~6.79} & \cellcolor[rgb]{0.93,0.93,0.93} \textcolor[rgb]{0.00,0.00,0.00}{~~7.75} & \cellcolor[rgb]{0.94,0.94,0.94} \textcolor[rgb]{0.00,0.00,0.00}{~~7.10} \\
E2 & \cellcolor[rgb]{0.94,0.94,0.94} \textcolor[rgb]{0.00,0.00,0.00}{~~6.50} & \cellcolor[rgb]{0.97,0.97,0.97} \textcolor[rgb]{0.00,0.00,0.00}{~~3.57} & \cellcolor[rgb]{0.92,0.92,0.92} \textcolor[rgb]{0.00,0.00,0.00}{~~9.17} & \cellcolor[rgb]{0.92,0.92,0.92} \textcolor[rgb]{0.00,0.00,0.00}{~~9.33} & \cellcolor[rgb]{0.91,0.91,0.91} \textcolor[rgb]{0.00,0.00,0.00}{~~9.51} \\
E3 & \cellcolor[rgb]{0.80,0.80,0.80} \textcolor[rgb]{0.00,0.00,0.00}{22.35} & \cellcolor[rgb]{0.81,0.81,0.81} \textcolor[rgb]{0.00,0.00,0.00}{20.86} & \cellcolor[rgb]{1.00,1.00,1.00} \textcolor[rgb]{0.00,0.00,0.00}{~~0.11} & \cellcolor[rgb]{1.00,1.00,1.00} \textcolor[rgb]{0.00,0.00,0.00}{~~0.14} & \cellcolor[rgb]{0.86,0.86,0.86} \textcolor[rgb]{0.00,0.00,0.00}{15.26} \\
E4 & \cellcolor[rgb]{0.85,0.85,0.85} \textcolor[rgb]{0.00,0.00,0.00}{16.32} & \cellcolor[rgb]{0.86,0.86,0.86} \textcolor[rgb]{0.00,0.00,0.00}{15.32} & \cellcolor[rgb]{1.00,1.00,1.00} \textcolor[rgb]{0.00,0.00,0.00}{~~0.17} & \cellcolor[rgb]{1.00,1.00,1.00} \textcolor[rgb]{0.00,0.00,0.00}{~~0.11} & \cellcolor[rgb]{0.89,0.89,0.89} \textcolor[rgb]{0.00,0.00,0.00}{11.77} \\
% E12 (low) &  &  &  &  &  \\
% E34 (low) &  &  &  &  &  \\
\midrule
PartialEdit & \cellcolor[rgb]{0.96,0.96,0.96} \textcolor[rgb]{0.00,0.00,0.00}{~~4.07} & \cellcolor[rgb]{0.97,0.97,0.97} \textcolor[rgb]{0.00,0.00,0.00}{~~3.30} & \cellcolor[rgb]{1.00,1.00,1.00} \textcolor[rgb]{0.00,0.00,0.00}{~~0.18} & \cellcolor[rgb]{1.00,1.00,1.00} \textcolor[rgb]{0.00,0.00,0.00}{~~0.16} & \cellcolor[rgb]{0.98,0.98,0.98} \textcolor[rgb]{0.00,0.00,0.00}{~~2.77} \\
\bottomrule
\end{tabular}
}
% \vspace{3mm}
% \begin{minipage}{0.41\textwidth}
% \centering
% \begin{tabular}{rccccc}
% \toprule
% F1 & E1 & E2 & E3 & E4 & Pool\\
% \midrule
% E1 & 87.12 & 87.75 & 75.70 & 76.39 &  \\
% E2 & 82.04 & 88.83 & 65.63 & 68.82 &  \\
% E3 & 19.10 & 21.28 & 99.65 & 99.62 &  \\
% E4 & 29.71 & 31.65 & 99.53 & 99.72 &  \\
% % E12 (low)         &    &    &    &   & \\
% % E34 (low)         &    &    &    &   & \\
% Pool             &    &    &    &   & \\
% \bottomrule
% \end{tabular}
% \end{minipage}
\vspace{-2pt}
\end{table}

% Subsets in PartialEdit can be split to two groups considering the similarity of their applied technologies: (1) E1 and E2, (2) E3 and E4.
The results for localization are presented in Table \ref{tab:loc_cross_speechedit}. Similar to our findings in utterance-level spoof detection discussed in Section~\ref{ssec:utter}, models perform best when trained on data that match the test data. Their performance degrades when testing on data generated by unseen models. For example, 
VoiceCraft (E1) and SSR-Speech (E2) share similar technology, while both E3 and E4 are based on Audiobox. 
Models trained on data generated by E1 or E2 achieve lower EERs on those subsets but perform worse on audio generated by Audiobox (E3 and E4), and vice versa. In particular, models trained on E3 or E4 achieve very good EERs on E3 and E4, but they cannot generalize to E1 or E2. % tend to have poor generalizability: Despite achieving EERs below 1\% on E3 and E4, the same models obtain an EER exceeding 15\% on audio language model-based subsets (E1 and E2).
% , highlighting their lack of generalization. 
% It is understandable, because the edited parts are from more advanced speech editing models, as in E1 and E2, so it is harder to perform precise localization on these sets when trained on E3/E4. 
Furthermore, training on the entire PartialEdit dataset with E1-E4 pooled together (last row) achieves good performance across all test sets. While this result is not surprising, it reaffirms the conclusion we reached from Section \ref{sec:detection}: Diversity of training data matters. %for deepfake detection. Performance on each subsets generated by different speech editing model can be improved when they are pooled together for training in the last row.
% subsets from different speech editing models are pooled together for training, the model can efficiently handle the testing data they generate. 

% \vspace{-3mm}
\section{Discussion}
% \subsection{Is codec-processed unedited speech real?}
% \subsection{\hspace{-0pt}Should codec-processed unedited speech be real?}
% \subsection{Localization on codec-processed speech}
% \subsection{Cut-and paste vs. codec processing in Localization}
\subsection{Impact of post-processing step of PartialEdit curation} %on deepfake localization}
% 
% \subsection{Bona Fide or Spoof? Context-Unedited Codec-Processed Region}
\label{sec:codec_real_fake}

% \nz{CodecFake~\cite{wu24p_interspeech} focuses on deepfake generated by neural audio tokenizers.
% % utilize codec-processed speech as a novel and challenging deepfake category. 
% We take CodecFake-E (the Encodec subset of CodecFake) as all our speech editing methods are based on the Encodec tokenizer. }\NZ{Discarded from 3}

As introduced in Section~\ref{sec:data_creation}, PartialEdit applies an additional cut-and-paste post-processing step on top of PartialEdit-Codec to mitigate artifacts introduced by neural codecs on content-unedited regions.
The key difference between PartialEdit-Codec and PartialEdit, therefore, is whether the content-unedited regions are processed by a neural codec or directly stitched from the original audio. 
Although results from Section~\ref{ssec:utter} indicate the superior performance when combining both stitching and codec artifacts compared to only including codec artifacts (i.e. PartialEdit vs. PartialEdit-Codec),
% artifacts introduced by post-processing with cutting and pasting are less detectable than those introduced by codec processing in the utterance-level deepfake detection, 
it remains unclear how those two operations affect the localization of edited regions in PartialEdit. This section conducts an experiment to examine deepfake localization on two settings in Table~\ref{tab:loc_infill_CutPaste}.  We include CodecFake~\cite{wu24p_interspeech} as it is the codec-processed version of VCTK, and we select the SSR-Speech-edited (E2) subset among PartialEdit, as it is one of the most recent approaches and is hard to detect according to Section~\ref{sec:localization}. To clarify, we define deepfake as content-edited regions, regardless of whether the segments have undergone codec processing, with the assumption that the detection target is malicious generation.

% Importantly, only the content-edited segments are labeled as fake—regardless of whether they have been codec-processed.
% We select the SSR-Speech-edited (E2) subset as spoof, as it is one of the most recent approaches and is hard to detect according to Section~\ref{sec:localization}. 
% Both VCTK and CodecFake-Encodec are considered independently as bona fide to correspond to content-unedited parts within PartialEdit and PartialEdit-Codec. 

% By considering content-unedited codec-processed regions as real in PartialEdit-Codec, 

\begin{table}[!t]
    \centering % \scriptsize
    \tabcolsep=0.17cm
    \caption{Comparison of localization EER (\%) on PartialEdit-Codec with different settings. $\triangle$ indicates datasets used as bona fide, while $\bigcirc$ represents datasets used as deepfake. (CFE: CodecFake~\cite{wu24p_interspeech}-Encodec; PEC: PartialEdit-Codec)}
    \label{tab:loc_infill_CutPaste}
\scalebox{0.90}{
    \begin{tabular}{c cccc S[table-format=2.2] S[table-format=2.2]}
    \toprule
    \multicolumn{1}{c}{\multirow{2}{*}{}} & \multicolumn{4}{c}{Train on}                                                                                & \multicolumn{2}{c}{Test on } \\ \cmidrule(lr){2-5} \cmidrule(lr){6-7}
\multicolumn{1}{c}{}                                                  & \multicolumn{1}{c}{VCTK} & \multicolumn{1}{c}{CFE} & \multicolumn{1}{c}{PEC} & \multicolumn{1}{c}{PartialEdit} & \multicolumn{1}{c}{I} 
 & \multicolumn{1}{c}{II}                               \\
    \midrule
    % I     & $\triangle$  &       &       & $\bigcirc$   & 3.57 & 47.14  \\
    % % II    & $\triangle$  &       & $\bigcirc$   &       & 17.04 & 18.33  \\
    % II   &       & $\triangle$  & $\bigcirc$   &       & 10.73& 5.30   \\
    I &  $\triangle$ &  &  &  $\bigcirc$ & \cellcolor[rgb]{0.97,0.97,0.97} \textcolor[rgb]{0.00,0.00,0.00}{~~3.57} & \cellcolor[rgb]{0.58,0.58,0.58} \textcolor[rgb]{0.00,0.00,0.00}{47.14} \\
% II &  $\triangle$ &  &  $\bigcirc$ &  & \cellcolor[rgb]{0.85,0.85,0.85} \textcolor[rgb]{0.00,0.00,0.00}{17.04} & \cellcolor[rgb]{0.84,0.84,0.84} \textcolor[rgb]{0.00,0.00,0.00}{18.33} \\
II &  &  $\triangle$ &  $\bigcirc$ &  & \cellcolor[rgb]{0.90,0.90,0.90} \textcolor[rgb]{0.00,0.00,0.00}{10.73} & \cellcolor[rgb]{0.95,0.95,0.95} \textcolor[rgb]{0.00,0.00,0.00}{~~5.30} \\
    \bottomrule
    % \multicolumn{6}{l}{$\triangle$: Dataset used as bona fide, $\bigcirc$: Dataset used as spoof.} 
    \end{tabular}%
}
\vspace{-2pt}
\end{table}

% 4x4
% \begin{table}[!t]
%     \centering % \scriptsize
%     \tabcolsep=0.17cm
%     \caption{Comparison of localization EER (\%) on PartialEdit-Codec with different training settings. $\triangle$ indicates datasets used as bona fide, while $\bigcirc$ represents datasets used as deepfake. (CFE: CodecFake-Encodec; PEC: PartialEdit-Codec)}
%     \label{tab:loc_infill_CutPaste}
% \scalebox{0.90}{
%     \begin{tabular}{c cccc S[table-format=2.2] S[table-format=2.2] S[table-format=2.2] S[table-format=2.2]}
%     \toprule
%     \multirow{2}{*}{Exp.} & \multicolumn{4}{c}{Datasets}  & \multicolumn{4}{c}{EER (\%) Performance} \\
%     \cmidrule(lr){2-5} \cmidrule(lr){6-9}
%           & VCTK  & CFE & PEC & PE & I  & II  & III  & IV  \\
%     \midrule
%     I     & $\triangle$  &       &       &    $\bigcirc$   & 3.57  & 21.10  & 47.14  & 35.81  \\
%     II    & $\triangle$  &       &    $\bigcirc$   &       & 17.04  & 5.71   & 18.33  & 13.32  \\
%     III   &       & $\triangle$  &     $\bigcirc$  &       & 10.73  & 10.78  & 5.30   & 8.37   \\
%     IV    & $\triangle$  & $\triangle$  &   $\bigcirc$    &      & 13.18  & 5.62   & 5.43   & 5.55   \\
% \bottomrule
% \end{tabular}
% }
% \end{table}

The results are presented in Table~\ref{tab:loc_infill_CutPaste}. As expected, the model can easily locate the content-edited region when the bona fide subset in the training data matches the content-unedited region in the test data. In such cases, the model achieves EERs lower than 6\% on the diagonal. 
Notably, we observe a high EER close to 50\% when training on I and testing on II, indicating that if content-unedited codec-processed regions are not seen in partially edited audio during training, it becomes difficult to accurately locate the content-edited regions surrounded by content-unedited codec-processed segments. This suggests that artifacts introduced by codec processing may mislead the model if they are not recognized as bona fide during training. Crucially, when including Encodec-resynthesized data as a bona fide subset in row II, the result improves when testing on I, suggesting an approach to mitigate the misleading effects of these artifacts.

\subsection{Limitations}

% As an initial study on partial deepfake audio in speech editing, 
We acknowledge a few limitations in this study.
% \textbf{on diverse codec types} Only Encodec is included, probably due to its easy-to-train and open-source. We cannot find speech editing models based on other codec. [What is the consequence]
\textbf{1) On diverse speech editing models}. Although we discussed more speech editing models compared with \cite{huang2024detecting}, this is still limited. With the advancement of neural audio codecs and audio language models, further analysis of more sophisticated speech editing models is worth exploring.
% \textbf{On text edit restrictions}.
% We asked the GPT to only modify one word. However, to maintain sentence naturalness sometimes two words have to be modified, for example ``included in'' to ``excluded from''. The time span for those two words can sometimes may be not continuous. 
\textbf{2) On variations in text editing}.
The localization task in our study focuses solely on identifying substitution regions. It does not address the localization of deletion operations, as detecting deletion requires further methodological design. However, SSR-Speech is capable of providing deletion or addition, though not in our PartialEdit-E2 (generated by SSR-Speech), which could be explored in future work.

% , as how to define locating deletions requires further discussion.
% , which poses unique challenges to partial deepfake detection. 
% Nevertheless, PartialEdit-E2 (generated by SSR-speech) provide deletion locations, which could be explored in future investigations.

% substitution / addition

% not deletion

\section{Conclusion}
% Main findings.

% Future Work: infilling/cut-and-paste methods, multilingual setting

In this study, we introduced PartialEdit, a partially edited deepfake dataset tailored for speech deepfake detection against neural speech editing. Unlike traditional deepfake datasets, PartialEdit consists of speech utterances where segments are modified by advanced speech editing algorithms and seamlessly stitched back into the original recording. Additionally, we include PartialEdit-Codec, where the unedited regions are also processed through a neural codec, reflecting the common operations of modern speech editing models.

Using PartialEdit, we investigated both deepfake detection and deepfake localization tasks. Our experiments reveal that models trained on PartialSpoof struggle to detect partially edited speech generated by neural speech editing models. Notably, among all models, VoiceCraft and SSR-Speech present greater challenges for detection.
% whereas Audiobox-generated speech is relatively easier to identify.

Furthermore, we clearly define bona fide and deepfake segments in partial deepfake localization: Deepfake segments should only refer to those whose content has been modified, while bona fide segments refer to content-unedited regions, regardless of whether they undergo codec processing. This definition respects the original intention of neural codec models---achieving more effective compression. 
% Our experiments show that this definition makes partial deepfake localization more challenging, 
% but the performance can be improved when codec-processed but content-unedited utterances are included as bona fide data in training.
% Our experiments show that to locate the content-edited region in partially edited speech, performance can be improved when codec-processed but content-unedited utterances are included as bona fide data in training.
Our experiments also show that including codec-processed but content-unedited utterances as bona fide examples during training can improve performance in localizing content-edited regions in partial deepfakes.
\vfill\pagebreak

\section{Acknowledgments}

% This material is based on work supported by the Audiobox License, Copyright © Meta Platforms, Inc. All Rights Reserved.

This work is supported in part by National Institute of Justice Graduate Research Fellowship Award 15PNIJ-23-GG-01933-RESS, Intelligence Advanced Research Projects Activity ARTS Program, a New York State Center of Excellence in Data Science Award, and Meta Audiobox Responsible Generation Grant. The GPU resources are supported by NAIRR Pilot Project \#240152, ACCESS \#ELE240019, and NCSA Delta. 

\noindent Part of this material is based on work supported by Audiobox License, Copyright\hspace{-1pt} © Meta Platforms, Inc.\hspace{-1pt} All Rights Reserved.

The authors would also like to thank Puyuan Peng (UT Austin) and Helin Wang (JHU) for their brief discussions and for open-sourcing their work on VoiceCraft and SSR-Speech.

% \ifinterspeechfinal
%      The Interspeech 2025 organisers
% \else
%      The authors
% \fi
% would like to thank ISCA and the organising committees of past Interspeech conferences for their help and for kindly providing the previous version of this template.

\bibliographystyle{IEEEtran}
\bibliography{mybib}

\end{document}